# Reconstructing Images of Two Adjacent Objects through Scattering Medium Using Generative Adversarial Network

**Xuetian Lai,**[a] **Qiongyao Li,**[a] **Ziyang Chen,**[a] **Xiaopeng Shao** [b,c*]**, Jixiong Pu**[a,c**]

[a]Fujian Provincial Key Laboratory of Light Propagation and Transformation, College of Information Science and Engineering, Huaqiao University, Xiamen, Fujian 361021, China
[b]School of Physics and Optoelectronic Engineering, Xidian University, Xi'an 710071, China
[c]Hangzhou Institute of Technology, Xidian University, Xiaoshan, Hangzhou 311200, China

**Abstract**. Reconstruction of image by using convolutional neural networks (CNNs) has been vigorously studied in the last decade. Until now, there have being developed several techniques for imaging of a single object through scattering medium by using neural networks, however how to reconstruct images of more than one object simultaneously seems hard to realize. In this paper, we demonstrate an approach by using generative adversarial network (GAN) to reconstruct images of two adjacent objects through scattering media. We construct an imaging system for imaging of two adjacent objects behind the scattering media. In general, as the light field of two adjacent object images pass through the scattering slab, a speckle pattern is obtained. The designed adversarial network, which is called as YGAN, is employed to reconstruct the images simultaneously. It is shown that based on the trained YGAN, we can reconstruct images of two adjacent objects from one speckle pattern with high fidelity. In addition, we study the influence of the object image types, and the distance between the two adjacent objects on the fidelity of the reconstructed images. Moreover even if another scattering medium is inserted between the two objects, we can also reconstruct the images of two objects from a speckle with high quality. The technique presented in this work can be used for applications in areas of medical image analysis, such as medical image classification, segmentation, and studies of multi-object scattering imaging *etc*.

\* Email: xpshao@xidian.edu.cn

\*\* E-mail: jixiong@hqu.edu.cn

## 1 Introduction

Imaging through scattering media remains one of the most important as well as challenging topics in computational optics. The light scattering process is complex as passing through a scattering medium. There are many optical paths for the light field propagation because of the microstructure of the scattering medium. The output light field is deteriorated and a speckle pattern instead of the original object image is obtained at the detector[1,2]. In order to invert the scattering process and reconstruct the original object image, various methods have been proposed in the past decade, such as phase conjugation,[3,4] wavefront shaping,[5] transmission matrix (TM) based method,[6,7] memory effect based method[8], and intensity correlation,[9-11] *etc*. These methods have been made major progress for realizing imaging through scattering media. However these methods are found to be of limitation. For examples, TM based method is of high susceptibility to external changes of the experiments and need complex device configuration; the memory effect based method is limited by the memory effect range,[12] *etc*.

Recently, deep learning based method has become a new trend in scattering imaging. The deep learning technique is found to be robust, and effective in imaging through scattering medium, due to that it does not need prior knowledge of the optical system and it simplifies the experimental configuration.[13-15] Especially, the convolutional neural networks (CNNs) show the state-of-the-art performance in reconstructing images from speckles.[16-23] The CNNs reconstruct images from speckles through end-to-end training, predicting a category label at each pixel in the input speckle. One of the most popular model used for scattering imaging is U-Net,[24] which shows great capability of extracting statistical features from input speckles, and the imaging quality is highly improved.

However, most of the studies of imaging through scattering media are mainly focused on the reconstruction of one object image, in which the object information can be extracted from a single speckle by using CNNs. How to reconstruct images of more than one object from a single speckle seems more important, and more challenging, due to that the information of the object images are highly mixed in the light propagation process. In addition, it is well known that for imaging from the speckle, most of the CNNs are trained to make pixel-wise predictions, treating each pixel of the image independently. Thus, it may lead to the lack of spatial continuity in the final result.[25,26] For getting the better performance, the network should not only focuses on characterizing the category of each pixel value, but also consider reinforcing the visual appearance of the reconstructed images.

In this paper, we propose an approach of using generative adversarial network (GAN) for imaging through strong scattering media. The GANs have been found remarkable applications in modeling image distributions,[27-30] and have been successfully employed to solve inverse problems such as image super-resolution and in-painting.[31,32] The proposed GAN consists of an imaging network (the generator) and a full-convolutional network (the discriminator). Specifically, the generator is a Y-Net,[33,34] thus we term the network as ***YGAN***. The two networks (i.e., the generator and the discriminator) aim at obstructing the target goal of each other in the training process. In this way, the loss function of the generator is optimized by the combination of reconstruction loss and adversarial loss,[35] which drives the generator to further modeling the distribution of Ground Truth. Thus the reconstruction accuracy and spatial continuity of the reconstructed images are improved.

*For the first time, we employ the proposed YGAN to realize simultaneous imaging of two different objects located on adjacent layers.* We demonstrate the performance of the YGAN with a series of experiments on an optical system. The trained YGAN can reconstruct the images of two different objects from a single speckle, and it is robust against distance change between the two objects. Moreover even if an additional scattering medium is inserted between the two objects, the trained YGAN can also reconstruct the images of the two objects from a single speckle with high accuracy.

## 2 Method

### 2.1 Network Architecture

The network architecture is shown in Figure1. The generator network is a full convolutional Y-Net, as illustrated in Figure 1(a). It follows the symmetrical skip-connection[36] of U-Net. Specifically, the two up-sampling paths share the generic feature representations extracted by the down-sampling path through skip-connection. The input is a single speckle and the outputs are two prediction maps of two object images. Because Max-pooling will cause information loss,[37] the convolution with stride of 2 is used in the down-sampling path until the $2\times2\times512$ activation maps. Each up-sampling path contains five blocks, in which each block consists of up-sampling, 3×3 convolution, leaky rectified linear unit (leaky Relu), batch normalization (BN), and concatenate. The output prediction is obtained after taking an additional up-sampling and a convolution.

The discriminator network is shown in Figure 1(b). The outputs of generator are concatenated to form partly the input of the discriminator. Also, the corresponding Ground Truth are concatenated and fed into the discriminator. In the network, the input size and the output size are 256×256×2, and 16×16, respectively, following the method of PatchGAN.[37] It is shown that, this strategy restricts the discriminator attention to patches, and the generator will produce image with high structural fidelity.

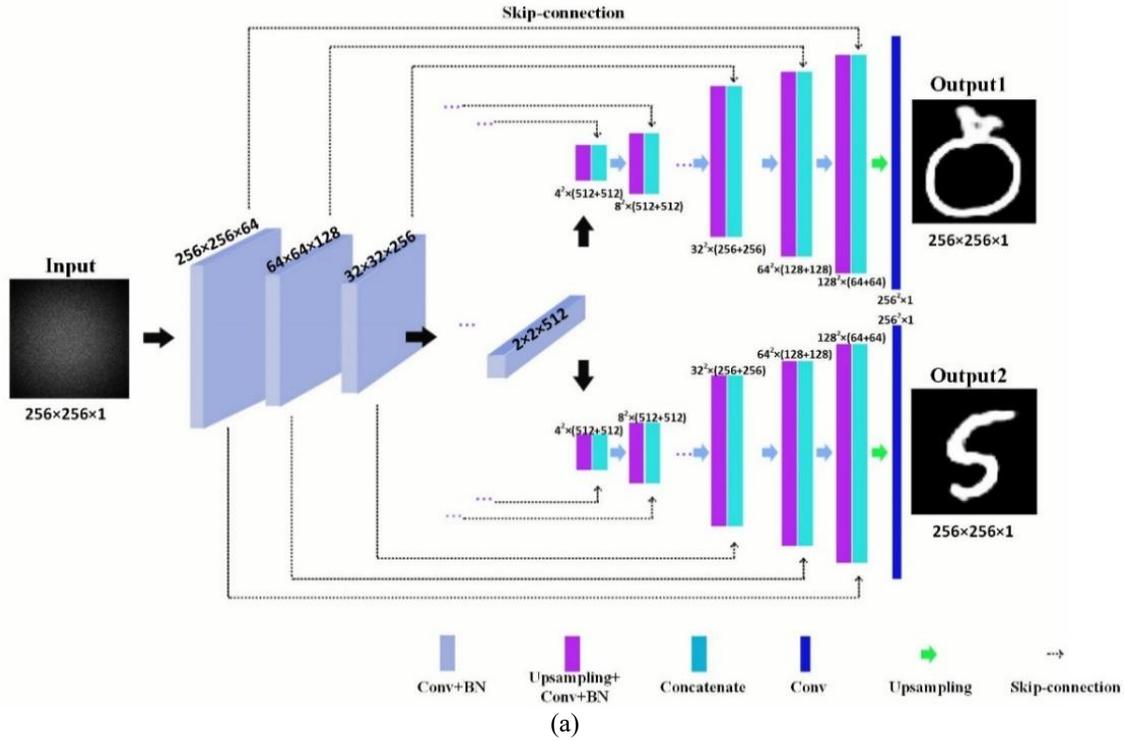

(a)

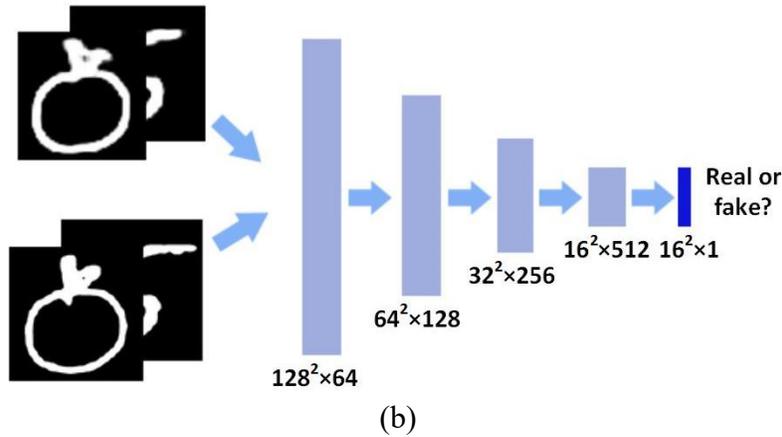

(b)

**Figure 1.** The architecture of YGAN. (a) The architecture of the generator network, which is a Y-type network. It takes a single speckle as input and outputs two prediction maps of the objects; (b) the architecture of the discriminator network, in which the inputs are partly from generator and partly from original images.

## 2.2 Objective Functions

The objective function is a hybrid loss function that is a weighted sum of two terms. We use $G(\cdot)$ to denote the outputs of the generator, and use $G_{o_1 o_2}(\cdot)$ to denote the output of concatenation, where $o_1$ and $o_2$ are the predicted outputs of the generator. Given an input speckle $x$ and corresponding Ground Truth images $y_1$ and $y_2$, the objective function of YGAN can be expressed as:

$$l(\theta_g, \theta_d) = \lambda[l(G(x), y_1) + l(G(x), y_2)] - [l(D(Y_{y_1 y_2}), 1) + l(D(G_{o_1 o_2}(x)), 0)], \quad (1)$$

where $\theta_g$ and $\theta_d$ denote parameters of the generator network and the discriminator network respectively. $Y_{y_1 y_2}$ denotes the concatenated image of $y_1$ and $y_2$. During training, the generator and the discriminator network are trained alternatively. The generator aims at minimizing the objective function, while the discriminator is trained to maximize it.

When training the discriminator, the second term of Eq.(1) will be minimized, which is equivalent to minimizing the following binary classification loss:

$$l(D(Y_{y_1 y_2}), 1) + l(D(G_{o_1 o_2}(x)), 0) \quad (2)$$

Following this, the discriminator is trained to predict that the images from the generator network are fake with label 0 and images from Ground Truth are real with label 1. For updating the discriminator, $l$ can be a binary-cross-entropy or other measurements, here, we use mean square error (MSE), which is formulated as:

$$MSE = \frac{1}{H*W} \sum_{i=1}^{H} \sum_{j=1}^{W} (\hat{y}(i,j) - y(i,j)), \quad (3)$$

where, $H$ and $W$ indicate the width and height of the image. $\hat{y}$ and $y$ refer to the predicted object image and the Ground Truth, respectively.

With the fixed discriminator network, the generator network will be trained to minimize the following terms:

$$\lambda[l(G(x), y_1) + l(G(x), y_2)] - l(D(G_{o_1 o_2}(x)), 0) \quad (4)$$

The first term is generally used in CNN for scattering imaging (see e.g. [11-14]), which drives a pixel-level mapping. The second term is based on the discriminator network, which can be considered as an auxiliary loss term to the generator network, penalizing the generator for producing blurry images. By minimizing the second term, the generator network is encouraged to produce images that are close to the distributions of the Ground Truth. In practice, we replace the second term by $+l(D(G_{o_1 o_2}(x)), 1)$[4]. For updating the generator network, we set $l$ to be binary-cross-entropy for reconstructing binary object images, which can be expressed as:

$$l(\hat{y}, y) = -y \ln \hat{y} + (1 - y) \ln(1 - \hat{y}). \quad (5)$$

For reconstructing grayscale object images, $l$ is MSE.

## 3 Experimental Setup and Data Requirement

### 3.1 Experimental Setup

The experimental setup is shown in Figure 2. The laser (MGL-F-532.8nm-2W) is transmitted through a micro-scope objective (OBJ$_1$, ×20, NA=0.25) and a pinhole aperture (D= 20μm). Next, the light field is collimated by Lens$_1$ (f = 20mm) and passes through a horizontal polarizer to illuminate SLM$_1$ (Holoeye, LC2012) and SLM$_2$ (Holoeye, LC2012, reflective) one after the other. The two SLMs are programmed to load object images one-by-one simultaneously. The distance between the two SLMs is chosen to be 45cm, 55cm respectively. The light field with overlapped information of two objects then passes through Lens$_2$ (f = 100mm) and a scattering slab S$_2$. The output scattered light is collected by OBJ$_2$ (×20, NA=0.25) and captured by CCD (AVT PIKE F-421B).

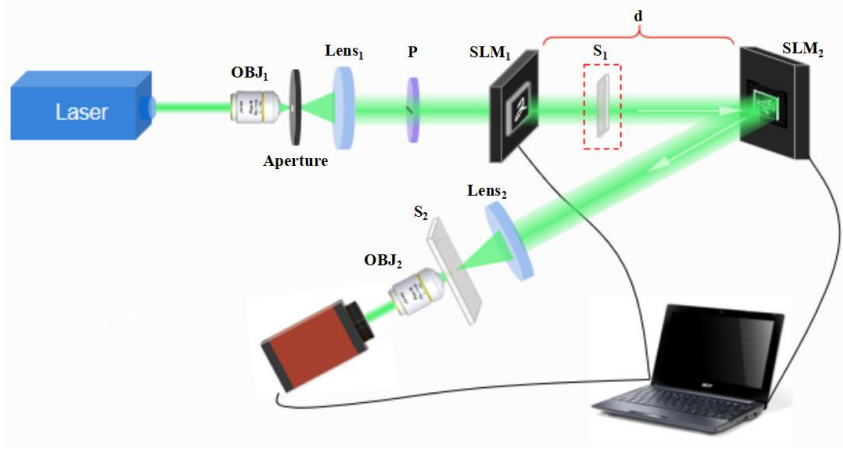

**Figure 2.** The optical system for collecting speckle datasets. $OBJ_1$ (×20, NA=0.25) and $OBJ_2$ (×20, NA=0.25) are micro-scope objective, $Lens_1$ (f = 20mm) and $Lens_2$ (f = 100mm) are used to collimate and focus the light field. P is a horizontal polarizer. $SLM_1$ (Holoeye, LC2012) and $SLM_2$ (Holoeye, LC2012, reflective) are used to display object images. The distance between $SLM_1$ and $SLM_2$ is chosen to be 45cm or 55cm. Both $S_1$ and $S_2$ are scattering medium, in which $S_1$ can be selectively inserted.

*3.2 Data Requirement*

The object images loaded onto the two SLMs are acquired from MNIST handwritten digits,[39] Quickdraw Objects,[40] and Fashion-mnist.[41] The handwritten digits and Quickdraw images are used as binary objects in this work. We randomly select 10000 images from [39] and [40], respectively. They are resized to 512*512 pixels and binarized to 0, 1 before loaded onto the two SLMs. The images of Fashion-mnist are used as grayscale objects. We randomly select 20000 grayscale images from Fashion-mnist and resize these images to 512*512 pixels. The first 10000 images are displayed on $SLM_1$ and the remaining 10000 images are displayed on $SLM_2$.

By using these object images, we purposely design three experiments for collecting the following speckle datasets:

*Group A:* we use the scattering slab with thickness of 200μm as $S_2$ ($S_1$ has not been set in this experiment yet), and the distance between the two objects is 45cm. The objects loaded on $SLM_1$ and $SLM_2$ are digits and Quickdraw images respectively. We collect 10000 speckles corresponding to the loaded objects.

*Group B1, B2:* we maintain the configuration of the above experiment and change the distance between the two objects (the distances are 45cm and 55cm, respectively). The objects loaded on $SLM_1$ and $SLM_2$ are grayscale images from Fashion-mnist. For each distance case, we collect 10000 speckles corresponding to the loaded objects.

*Group C:* we insert an additional scattering media $S_1$ (a diffuser, 600-grit) between $SLM_1$ and $SLM_2$, and displace $S_2$ by a diffuser (2000-grit). The objects loaded on $SLM_1$ and $SLM_2$ are digits and Quickdraw images respectively. We collect 10000 speckles corresponding to the loaded objects.

All the collected datasets are randomly divided into training and testing sets, in which 9000 speckle-object pairs belong to training set and 1000 speckle-object pairs belong to testing set. The testing set is only used to evaluate the performance of trained networks, which is not used during training. Before fed into the network, the speckle-object pairs of training set are downsampled from 512×512 pixels to 256×256 pixels and normalized between 0, 1, for reducing the network parameters and training time.

## 4 Results and Network Analysis

To show the advantages of our proposed YGAN, we compare the performance of our YGAN to other two networks without adversarial training: (**i**) the Y-Net, and (**ii**) the Y-Net1. The network Y-Net has the same architecture as the generator network. The network Y-Net1 is the modified Y-Net, in which the convolution stride is 1, and max-pooling layers are added for down-sampling.

### 4.1 Reconstructing images of two binary objects

First of all, we employ three networks to reconstruct binary object images from a single speckle. For the adversarial training of the YGAN, we alternately train the discriminator and generator step by step. The generator network is trained by using a loss weight $\lambda=100$ and learning rate $10^{-4}$, and both generator and discriminator use mini-batch SGD and Adam with momentum parameter $\beta=0.5$. The last two layers of generator use sigmoid as activation function and the related loss function of binary cross-entropy. We choose the hyper-parameters of the Y-Net and the Y-Net1, which are the same as those of the YGAN. These hyper-parameters are training epoch, learning rate, optimizer, activation functions and loss functions etc.

We train the three networks with the training set of ***Group A***. After training, the network performances are tested by speckles from the testing set. Examples of the test results of all the three networks are shown in Figure 3, demonstrating that the three networks are able to reconstruct two adjacent binary objects from a single speckle. Meanwhile, the images reconstructed by YGAN are visually closer to the real images than those by the other two networks, due to that YGAN uses adversarial training. In addition, both the reconstruction quality of YGAN and Y-Net are better than that of Y-Net1, indicating that it is feasible to use Y-type network with a stride of 2 in the encode path of the network.

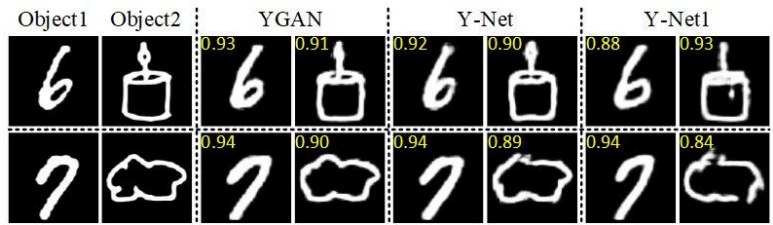

**Figure 3.** Reconstruction results of images of two binary objects on the adjacent layer by using the YGAN, Y-Net, and Y-Net1, respectively. The digits and Quickdraw images are reconstructed simultaneously.

To quantitatively analyze the performance of the three networks, we use the Structure Similarity (SSIM)[42] index to measure the similarity between the reconstructed object images and Ground Truth. The SSIM is formulated as:

$$SSIM = \frac{(2\mu_y\mu_{\hat{y}}+c_1)(2\sigma_{y\hat{y}}+c_2)}{(\mu_y^2+\mu_{\hat{y}}^2+c_1)(\sigma_y^2+\sigma_{\hat{y}}^2+c_2)}, \quad (6)$$

where, $y$ and $\hat{y}$ are Ground Truth and the reconstructed image, respectively; $\mu_y$, $\mu_{\hat{y}}$ refer to the means; $\sigma_y$ and $\sigma_{\hat{y}}$ refer to the variance; $\sigma_{y\hat{y}}$ refer to covariance of $y$ and $\hat{y}$; $c_1$ and $c_2$ are constants.

The averaged SSIM indices computed over all the test results of the networks are shown in Table 1. The averaged SSIM of YGAN is higher than that of Y-Net and Y-Net1, which illustrates the potential of the adversarial training on reversing scattering. The validating accuracy curves as a function of the training iteration step are shown in Figure 4. As the training iteration step grows, the networks learn to extract the invariant features of two original object images from a single speckle, and the validating accuracy gradually converge. At the same time, the YGAN and Y-Net converge to high values. However the Y-Net1 converges to a lower value.

**Table 1** Averaged SSIM of all the test results of YGAN, Y-Net, and Y-Net1, respectively.

|  | YGAN | Y-Net | Y-Net1 |
|---|---|---|---|
| **Digit** | 0.9240 | 0.9236 | 0.9010 |
| **Quickdraw** | 0.8721 | 0.8699 | 0.8128 |

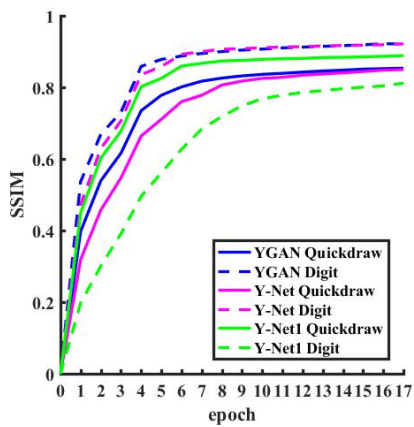

**Figure 4.** The validating accuracy curves for reconstructing images of two binary objects during training when using YGAN (blue), Y-Net (magenta), and Y_Net1 (green), respectively.

### 4.2 Reconstructing images of two grayscale objects

Next, the three networks are used to reconstruct two different grayscale object images from a single speckle. The networks are trained on the training set of ***Group B1***, ***B2*** with $L_1$ loss, which encourages less blurring.[26] The test results of the trained networks are shown in Figure 5. As can be seen from Figure 5, YGAN obtains the best quality of the reconstructed grayscale images among the three networks. The reconstruction results of YGAN retain most of the structural information of the original grayscale images, while that of Y-Net and Y-Net1 show ambiguity in the details of reconstructed images. Table 2 presents the averaged SSIM for all the test results of the three networks. Figure 6 illustrates the validating accuracy curves during training. The averaged SSIM of YGAN is 10% to 20% higher than that of the other two networks. Moreover, we observe that the generalization ability of YGAN continues to increase and converges to the values that are higher than that of the other two networks. These results further demonstrate the effectiveness of using the approach of adversarial training, showing the superior ability of YGAN on scattering imaging.

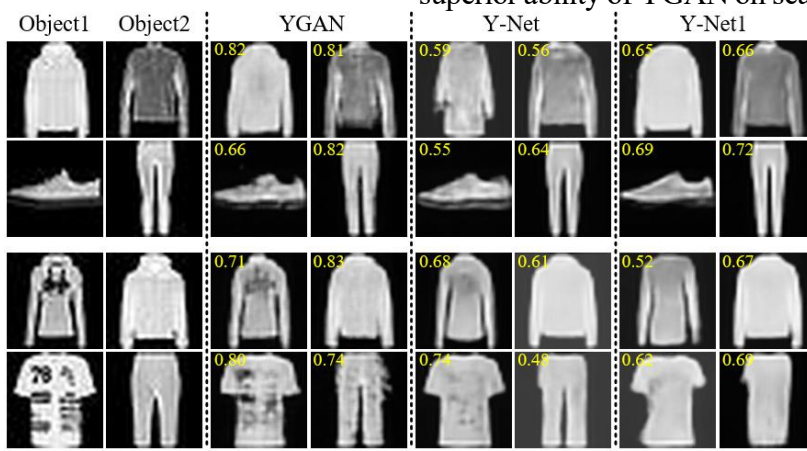

**Figure 5.** Reconstruction results of images of two grayscale objects on the adjacent layer by using the YGAN, Y-Net, and Y-Net1, respectively. The two rows on the top present the results of distance 45cm; the two rows on the bottom present the results of distance 55cm.

**Table 2** Averaged SSIM of all the test results of YGAN, Y-Net, and Y-Net1, respectively.

|  |  | YGAN | Y-Net | Y-Net1 |
|---|---|---|---|---|
| d = 45cm | Object 1 | 0.6869 | 0.5915 | 0.6313 |
|  | Object 2 | 0.7329 | 0.5400 | 0.6819 |
| d = 55cm | Object 1 | 0.6584 | 0.6555 | 0.5448 |
|  | Object 2 | 0.7002 | 0.5038 | 0.5794 |

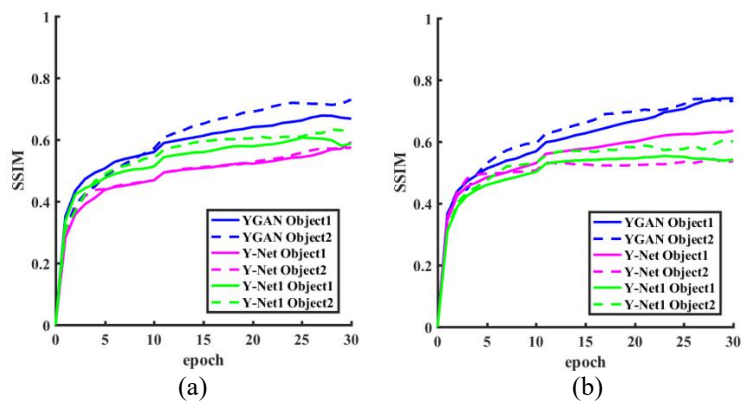

**Figure 6.** The validating accuracy curves for reconstructing images of two grayscale objects during training when using YGAN (blue), Y-Net (magenta), and Y_Net1 (green), respectively. (a) the curves in the case of d=45cm, (b) the curves in the case of d=55cm.

### 4.3 Reconstructing images of two objects in the case of two scattering media

Now we consider more complicated case in which there exist two scattering media. The first scattering medium ($S_1$) is located between the two objects ($SLM_1$ and $SLM_2$, as shown in Figure 2), and another scattering medium ($S_2$) is located in front of $Lens_2$. In this situation, we employ the YGAN for reconstructing images. To do this, we train the YGAN by the training set of *Group C*. It is shown from Figure 7 that the reconstructing accuracy of the results are as good as that of the reconstruction results in the experiment without the scattering medium $S_1$. This indicates that by using YGAN we can reconstruct image of more than one object with high accuracy, and the existing of the scattering medium $S_1$ between the two objects does not influence the reconstructing performance.

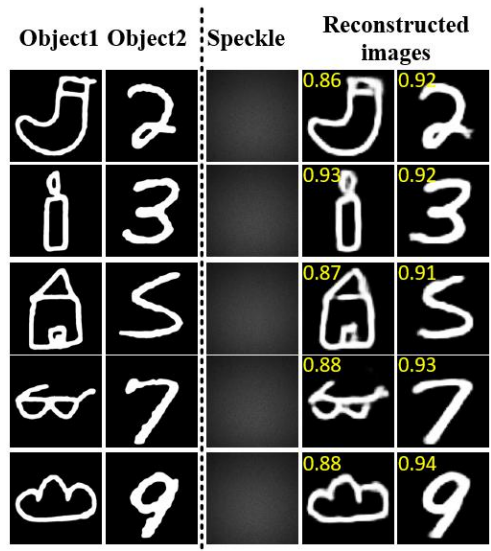

**Figure 7.** Reconstruction results of images of two objects on the adjacent layer by using the YGAN.

There exist two scattering media in the optical system, and the two images are reconstructed simultaneously.

## 5 Conclusion

We have proposed a neural network—YGAN to reconstruct images of two different objects on adjacent layers behind scattering media. It has been demonstrated that the YGAN can be used for reconstructing images of two binary objects from one speckle with high fidelity. Moreover, it has shown that the YGAN performs very well when used for reconstructing images of two grayscale objects. In addition, as the distance between the two grayscale objects is changed, we can still obtain high reconstruction quality by using YGAN. Furthermore, even if there exist two scattering media in the optical system, the trained YGAN can also reconstruct images of two objects with high accuracy. It has been demonstrated that the proposed YGAN possesses superior robustness over other networks (Y-Net and Y-Net1) in reconstructing images of two objects through scattering media. Our proposed technique may find applications in medical image analysis, such as medical image classification, segmentation, and studies of multi-object scattering imaging *etc*.